\documentclass[pra,aps,twocolumn]{revtex4}
\usepackage{bbm}
\usepackage{}
\usepackage{mathrsfs}
\usepackage{amsfonts}
\usepackage{txfonts}
\usepackage{amssymb}
\usepackage{graphicx}
\usepackage{bm}
\usepackage{color}

\newcommand{\ket}[1]{|#1\rangle}
\newcommand{\bra}[1]{\langle #1|}

\begin{document}
\title{Average value estimation in nonadiabatic holonomic quantum computation}
\author{Guo-Fu Xu}
\email{xgf@sdu.edu.cn}
\affiliation{Department of Physics, Shandong University, Jinan 250100, China}
\author{P. Z. Zhao}
\email{pzzhao@sdu.edu.cn}
\affiliation{Department of Physics, Shandong University, Jinan 250100, China}
\date{\today}

\begin{abstract}
Nonadiabatic holonomic quantum computation has been attracting continuous attention since it was proposed. Until now, various schemes of nonadiabatic holonomic quantum computation have been developed and many of them have been experimentally demonstrated. It is known that at the end of a computation, one usually needs to estimate the average value of an observable. However, computation errors severely disturb the final state of a computation, causing erroneous average value estimation. Thus for nonadiabatic holonomic quantum computation, an important topic is to investigate how to better give the average value of an observable under the condition of computation errors. While the above topic is important, the previous works in the field of nonadiabatic holonomic quantum computation pay woefully inadequate attention to it. In this paper, we show that rescaling the measurement results can better give the average value of an observable in nonadiabatic holonomic quantum computation when computation errors are considered. Particularly, we show that by rescaling the measurement results, $56.25\%$ of the computation errors can be reduced when using depolarizing noise model, a widely adopted noise model in quantum computation community, to analyse the benefit of our method.
\end{abstract}
\maketitle
\date{\today}

\section{Introduction}

Unlike classical computation, quantum computation can use quantum parallelism to process information encoded in physical systems. For this reason, quantum computation can solve many problems, such as factoring large integers and searching unsorted databases, much faster than classical computation \cite{Nielsen2001}. However, while the advantages of quantum computation are attracting, achieving them in practice is difficult. The main reason is that compared to classical systems, quantum systems are much easier to be affected by noise, so that quantum computation, which builds on quantum systems, is difficult to be realized with high fidelity. To overcome the noise problem and thereby realize high-fidelity quantum computation, researchers pay continuous attention to investigate robust quantum computation and until now impressive progresses have been made in this direction.

Geometric phases are important both in theory and application. The first kind of geometric phases discovered by researchers were adiabatic and Abelian geometric phases \cite{Berry}. This kind of geometric phases can be acquired by evolving a quantum system in a nondegenerate eigenstate adiabatically and cyclicly. Soon after, the notion of adiabatic and Abelian geometric phases was gradually generalized: a quantum system with degenerate eigenstates undergoing adiabatic cyclic evolution can acquire adiabatic and non-Abelian geometric phases or adiabatic quantum
holonomies \cite{Wilczek}; a quantum system with nondegenerate eigenstates undergoing nonadiabatic cyclic evolution can acquire nonadiabatic and Abelian geometric phases \cite{Aharonov}; a quantum system with degenerate
eigenstates undergoing nonadiabatic cyclic evolution can acquire nonadiabatic
and non-Abelian geometric phases or nonadiabatic quantum
holonomies \cite{Anandan}. Besides the above seminal works, there are also other remarkable works enriching the field of geometric phases \cite{Sjoqvist2000,Tong2004}.

Since geometric phases are only dependent on the path in which the quantum system evolves but independent of its evolutional details, quantum computations based on geometric phases are robust against certain control errors. As one important geometric quantum computation paradigm, nonadiabatic holonomic quantum computation \cite{Sjoqvist2012,Xu2012} builds its gates on nonadiabatic  and non-Abelian geometric phases \cite{Anandan}. Moreover, nonadiabatic holonomic quantum computation does not have the constraint of adiabatic evolution condition \cite{Tong2005,Tong2007,Tong2010} and thereby has the feature of being implemented with high-speed. Because of the above features, nonadiabatic holonomic quantum computation has been attracting continuous attention since it was proposed. Until now, a number of relevant schemes have been put forward \cite{Abdumalikov,Feng,Arroyo,Zu,Liang,Zhang1,Mousolou,Xue,Xu3,Sjoqvist2,
Herterich,Zhang2,Wang,Sun,Xue2017,Li2017,
Sekiguchi,Zhou2017,Hong2018,Zhao2019,danilin18,Xu2018,Zhang2019,Ramberg,Liu2019,
Zhu2019,Yan2019,Ai2020,Xu2021,Zhao2021,Liang2022,Shen2023,Zhang2023}, and some schemes have been experimentally demonstrated in circuit QED \cite{Abdumalikov,danilin18,Xu2018,Yan2019,Zhang2019}, nuclear magnetic resonance systems \cite{Feng,Li2017,Zhu2019}, nitrogen-vacancy centers \cite{Arroyo,Zu,Sekiguchi,Zhou2017} and trapped ions \cite{Ai2020}.

When using quantum computation to implement a computational task, an important step is to estimate the average value of an observable at the end. However, computation errors can disturb the final state of the computation, thereby affecting the estimation of the average value. When implementing a computational task, many nonadiabatic holonomic gates are needed. While these nonadiabatic holonomic gates have robustness, they can not be perfect in practice. And these imperfections can accumulate, resulting in severe computation errors. Thus for nonadiabatic holonomic quantum computation, it is of significance to investigate how to better give the average value of an observable when the above computation errors are taken into account.

In this paper, we show that when computation errors in nonadiabatic holonomic quantum computation are considered, rescaling the measurement results is a better way to give the average value of an observable than the conventional way. Our proposal is based on the fact that while the ideal final state of nonadiabatic holonomic quantum computation resides in the logical space, the support of the noisy final state can occupies the whole Hilbert space. We also use depolarizing noise model, which is a widely adopted noise model in quantum computation community, to conduct the analysis and find that $56.25\%$ of the computation errors can be reduced when using the rescaling method to give the average value.

\section{the framework}

We now start to illustrate our framework. Before proceeding further, we first briefly review how to realize a nonadiabatic holonomic gate. We consider an $N$-dimensional quantum system governed by Hamiltonian $H(t)$, of which the evolution operator is denoted as $U(t)=\mathbf{T}\exp[-i\int^t_0 H(t^\prime )dt^\prime]$ with $\mathbf{T}$ being time ordering. We use $\{\ket{\phi_\mu(t)}\}^N_{\mu=1}$ to represent $N$ orthonormal solutions of the Schr\"{o}dinger equation $i{\partial\ket{\phi_\mu(t)}}/{\partial{t}}=H(t)\ket{\phi_\mu(t)}$. Assume there is an $L$-dimensional subspace $\mathcal{S}(t)=\mathrm{Span}\{\ket{\phi_\mu(t)}\}^L_{\mu=1}$ evolving cyclicly with the period $\tau$, i.e. $\mathcal{S}(\tau)=\mathcal{S}(0)$, and satisfying the parallel transport condition, i.e.,  $\langle
\phi_\mu(t)|H(t)|\phi_\nu(t)\rangle=0$, $\mu,\nu=1,2,\cdots,L$. The computational basis can be then encoded into $\mathcal{S}(0)$ and the final evolution operator $U(\tau)$ acting on $\mathcal{S}(0)$ is a nonadiabatic holonomic gate.

From the above review, one can readily see that to realize a nonadiabatic holonomic gate, the logical space needs to be smaller than the whole Hilbert space, i.e., the logical space is just a subspace of the whole Hilbert space. Thus, instead of using two-level systems, one usually uses three-level systems to build nonadiabatic holonomic quantum computation and for each three-level system, only two of its three internal states are used as logical states \cite{Sjoqvist2012}.

Clearly, when using nonadiabatic holonomic quantum computation to implement a specific computational task, one needs more than one three-level systems, and without loss of generality, we assume the required number is $n$.
\begin{figure}[htb]
  \includegraphics[scale=0.3]{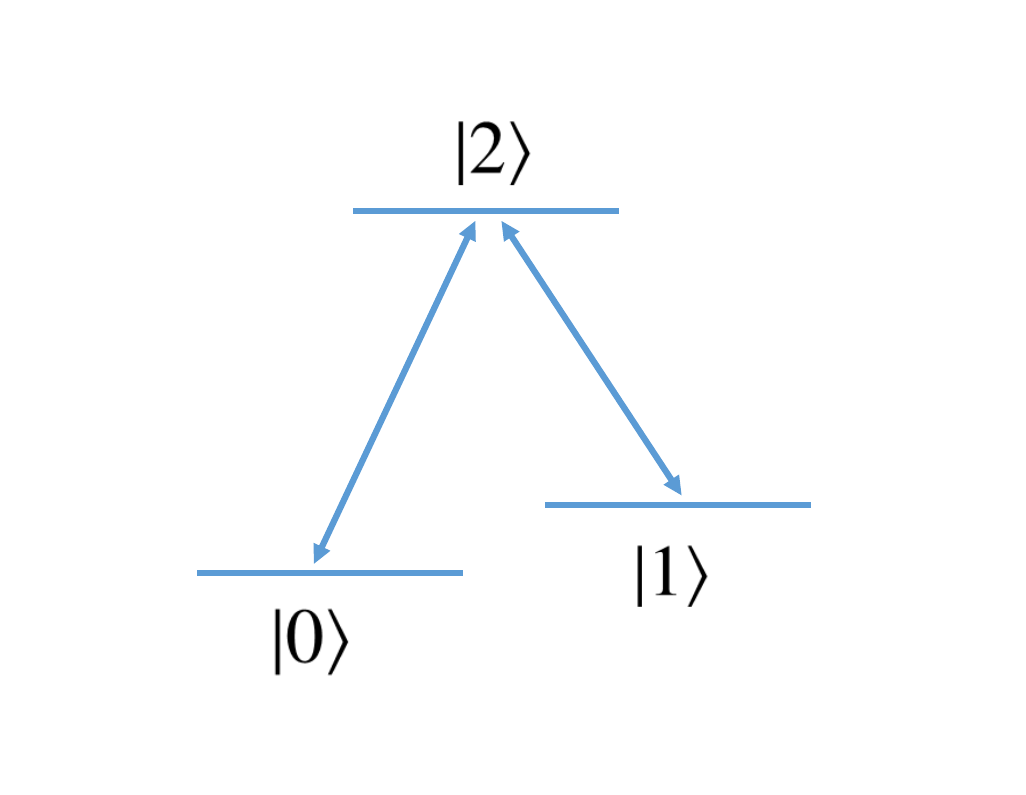}
  \caption{The structure of each of the $n$ three-level systems. The three states are denoted by $\ket{0}$, $\ket{1}$ and $\ket{2}$, and they form a $\Lambda$ structure.}
   \label{fig1}
\end{figure}
As shown in Fig.~\ref{fig1}, for each of these $n$ three-level systems, we denote its three states by $\ket{0}$, $\ket{1}$ and $\ket{2}$, respectively. Between these three states, the transitions $\ket{0}\leftrightarrow\ket{2}$ and $\ket{1}\leftrightarrow\ket{2}$ are allowed, while the transition $\ket{0}\leftrightarrow\ket{1}$ is forbidden. Of these three states, the states $\ket{0}$ and $\ket{1}$ are used as logical states and the state $\ket{2}$ is used as an auxiliary state. When implementing a computational task, these $n$ three-level systems are first prepared in an initial state $\rho$, i.e., the initial state of the computation. Then a family of nonadiabatic holonomic gates $\mathcal {G}_j$'s are performed on $\rho$, generating the final state $\rho_f$ of the computation. That is,
\begin{eqnarray}
\rho_f=\mathcal {G}_m\cdots\mathcal {G}_3\cdot\mathcal {G}_2\cdot\mathcal {G}_1(\rho),
\end{eqnarray}
where $m$ is the number of the performed nonadiabatic holonomic gates in the computation. At the end, a measurement is performed on the final state $\rho_f$, aiming to give the average value of some observable. That is,
\begin{eqnarray}
E=\text{Tr}(\rho_f{\hat{O}}), \label{e}
\end{eqnarray}
where $\hat{O}$ is the observable whose average value we want to estimate and $E$ denotes the average value. The above procedure can also be seen from Fig.~\ref{fig2}.
\begin{figure}[htb]
  \includegraphics[scale=0.25]{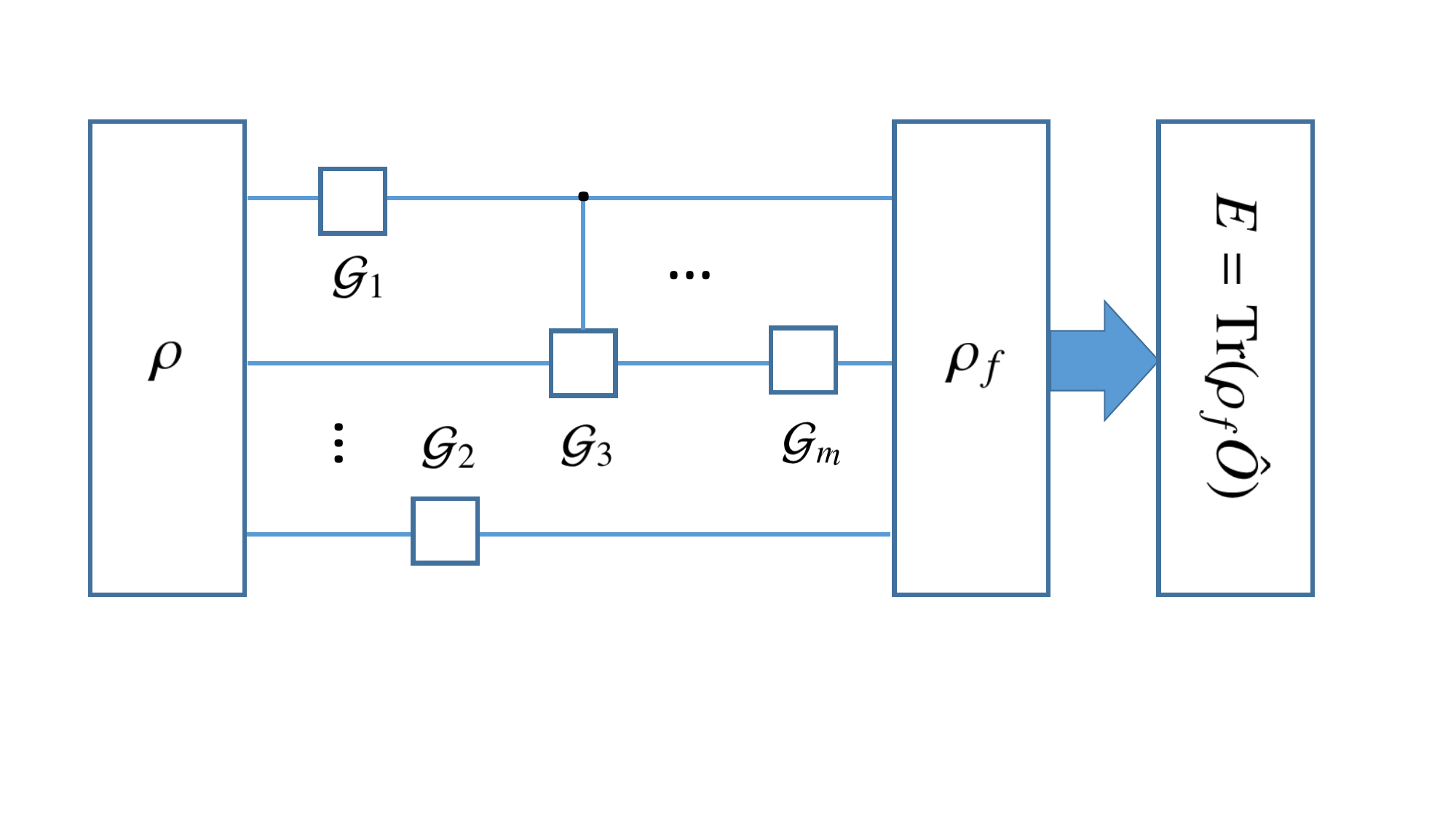}
  \caption{The procedure of using nonadiabatic holonomic quantum computation to implement a computational task. $\rho$ is the initial state of the computation, $\rho_f$ is the final state of the computation, $\mathcal {G}_j$'s are the nonadiabatic holonomic gates used in the computation, and $E=\text{Tr}(\rho_f{\hat{O}})$ is the average value we want to get.}
   \label{fig2}
\end{figure}

However, while nonadiabatic holonomic quantum gates have robustness, they can not be perfect in practice i.e., they can be noisy \cite{Solinas,Johansson}. Particularly, many nonadiabatic holonomic quantum gates are needed for implementing a computational task, and the imperfections of these gates can be accumulated, seriously affecting the quality of the final state $\rho_f$. Specifically, in practice we can not get the ideal final state $\rho_f$, but instead we get a final state written as
\begin{eqnarray}
\rho_f^\prime=\mathcal {G}_m^\prime\cdots
\mathcal {G}_3^\prime\cdot\mathcal {G}_2^\prime\cdot\mathcal {G}_1^\prime(\rho),
\end{eqnarray}
where $\mathcal {G}_j^\prime$ represents the $j$-th noisy nonadiabatic holonomic gate and $\rho_f^\prime$ represents the noisy final state of the computation. In this case, if the conventional way is used to estimate the average value of $\hat{O}$, one will get
\begin{eqnarray}
E^\prime=\text{Tr}(\rho_f^\prime{\hat{O}}), \label{eprime}
\end{eqnarray}
instead of the desired average value $E=\text{Tr}(\rho_f{\hat{O}})$.

Clearly, $E^\prime$ is not a good estimation of the desired average value $E=\text{Tr}(\rho_f{\hat{O}})$. To improve the estimation, we analyse the difference between the ideal final state $\rho_f$ and the noisy final state $\rho_f^\prime$. Recall that for each of the $n$ three-level systems, the states $\ket{0}$ and $\ket{1}$ are used to encode the logical information, while the state $\ket{2}$ is used as an auxiliary state. Thus, for these $n$ three-level systems, the whole Hilbert space is $\mathscr{H}=\{\ket{0},\ket{1},\ket{2}\}^{\otimes{n}}$, while the logical space is $\mathscr{L}=\{\ket{0},\ket{1}\}^{\otimes{n}}$. As it is well known, if a nonadiabatic holonomic gate is perfect, it transforms states in the logical space to states in the logical space. Thus, the support of the ideal final state $\rho_f$ is a subspace of the logical space $\mathscr{L}$. On the other hand, when the performed nonadiabatic holonomic gates $\mathcal {G}_j$'s are noisy, we do not expect the support of the noisy final state $\rho_f^\prime$ to be a subspace of the logical space $\mathscr{L}$ because the computation errors can cause the logical information to leak from the logical space. And the leakage problem can be induced by either the inaccuracy of the system Hamiltonian \cite{ Spiegelberg,Alves} or decoherence. Generally, the relation between $\rho_f^\prime$ and $\rho_f$ can be simply expressed as
\begin{eqnarray}
\rho_f^\prime=(1-P_\epsilon)\rho_f+P_\epsilon\rho_\epsilon, \label{relation}
\end{eqnarray}
where $P_\epsilon$ is a probability describing the strength of the computation errors and $\rho_\epsilon$ is a noisy state. Note that the support of $\rho_f$ is a subspace of the logical space $\mathscr{L}$, but the support of $\rho_\epsilon$ can be the whole Hilbert space $\mathscr{H}$. Thus, if we detect the state outside the logical space, we known errors have occurred. This inspires us to use the quantum error detection principle to reduce the errors \cite{Nielsen2001}. Specifically, based on the difference between $\rho_f$ and $\rho_\epsilon$, we consider the following projector
\begin{eqnarray}
\hat{P}=(\ket{0}\bra{0}+\ket{1}\bra{1})^{\otimes{n}}. \label{projector}
\end{eqnarray}
According to Eqs.~(\ref{relation}) and (\ref{projector}), one can see that the weight of the ideal final state $\rho_f$ within $\alpha\hat{P}\rho_f^\prime\hat{P}$ is higher than that within $\rho_f^\prime$, where $\alpha$ is a normalization factor. The reason for the above is that under the action of the projector $\hat{P}$, the ideal final state $\rho_f$ is totally retained, i.e., $\hat{P}\rho_f\hat{P}=\rho_f$, while the noisy state $\rho_\epsilon$ is only partly retained. The above discussion indicates that it is better to extract the information of the average value of $\hat{O}$ from $\alpha\hat{P}\rho_f^\prime\hat{P}$ than from $\rho_f^\prime$.

To proceed further, we analyse the properties of the observable $\hat{O}$. Because $\hat{O}$ is an observable, we can choose the eigenvectors of $\hat{O}$ so that these eigenvectors constitute an orthonormal basis for the whole Hilbert space $\mathscr{H}$. Without loss of generality, we denote the eigenvectors of the observable $\hat{O}$ by $\ket{j}$ and the eigenvalue corresponding to $\ket{j}$ by $\lambda_j$. As mentioned before, $\{\ket{j}\}$ constitute an orthonormal basis for the whole Hilbert space $\mathscr{H}$. Since the support of $\hat{O}$ is a subspace of the logical space $\mathscr{L}$, we can always appropriately choose $\{\ket{j}\}$ so that they can be divided into two parts: some of the eigenvectors are in the logical space $\mathscr{L}$ and the others are in the subspace $\mathscr{L}_\perp$, where $\mathscr{L}_\perp$ is the subspace orthogonal to the logical subspace. Then extracting the information of the average value of $\hat{O}$ from $\alpha\hat{P}\rho_f^\prime\hat{P}$ is equivalent to the following formula
\begin{eqnarray}
E_r=\frac{\sum_jP_j\lambda_j}{\sum_jP_j}~~~\text{s.t.}~~~\ket{j}
\in\mathscr{L}, \label{er}
\end{eqnarray}
where $P_j=\text{Tr}(\rho_f^\prime\ket{j}\bra{j})$ and by $\text{s.t.}~\ket{j}\in\mathscr{L}$, we mean the summation $\sum_j$ is only calculated for the eigenvectors belonging to the logical space $\mathscr{L}$. With the eigenvectors of $\hat{O}$ denoted by $\ket{j}$ and eigenvalues denoted by $\lambda_j$, we can also rewrite $E^\prime=\text{Tr}(\rho_f^\prime{\hat{O}})$ in Eq.~(\ref{eprime}) as follows
\begin{eqnarray}
E^\prime=\sum_jP_j\lambda_j. \label{eprime1}
\end{eqnarray}
According to Eqs.~(\ref{er}) and (\ref{eprime1}), one can readily see the difference: one is the summation range and the other is that the probabilities in Eq.~(\ref{er}) are rescaled by the factor $\sum_jP_j$ while the probabilities in Eq.~(\ref{eprime1}) are not rescaled.

In the above, we have shown that extracting the information of the average value of the observable $\hat{O}$ from $\alpha\hat{P}\rho_f^\prime\hat{P}$ is better than from $\rho_f^\prime$, that is, Eq.~(\ref{er}) is better to estimate the desired average value of the observable $\hat{O}$ than Eq.~(\ref{eprime1}). In the following, we will analyse to what extent one can get benefit from using the rescaling method, i.e., Eq.~(\ref{er}).

It is known that depolarizing noise model is widely used to describe computation errors in quantum computation community. Thus we here adopt this noise model to conduct our analysis. As shown in Fig.~\ref{fig2}, a family of nonadiabatic holonomic gates $\mathcal {G}_j$'s are used in the computation. Usually, these nonadiabatic holonomic gates are one-qubit and two-qubit gates. That is, only one-qubit and two-qubit gates are used to process the information. Moreover, these gates are not perfect but experience depolarizing noise \cite{Bhattacharyya}. Since the quality of the gates is high, it is reasonable to assume only one gate in the computation is erroneous. Because one-qubit gates are much more reliable than two-qubit gates, the erroneous gate in the computation can be assumed to be a two-qubit gate.

Without loss of generality, we assume the erroneous two-qubit gate acts on the three-level systems $a$ and $b$, that is,
\begin{eqnarray}
\mathcal {G}^\prime_{k}=\mathcal {N}_{ab}\cdot\mathcal {G}_{ab},
\end{eqnarray}
where $k\in\{1,2,\cdots,m\}$, $\mathcal {G}_{ab}=\mathcal {G}_{k}$ represents the ideal gate, and $\mathcal {N}_{ab}$ represents the errors. {\it It is very important to note that $k$ is not a fixed number.} Recall that we have assumed only one gate in the computation is erroneous. But this does not mean a fixed gate is erroneous every time we implement the computation. Instated, this means that every time we implement the computation, one of the performed gates is erroneous but which one is erroneous is not fixed.

$\mathcal {N}_{ab}$ has the possible values described by the generalized Pauli operators
\begin{eqnarray}
(X)^{a_1}(Z)^{a_2}\otimes(X)^{b_1}(Z)^{b_2}. \label{gp}
\end{eqnarray}
In the above, operators $(X)^{a_1}(Z)^{a_2}$ and $(X)^{b_1}(Z)^{b_2}$ respectively act on three-level systems $a$ and $b$, where $a_1,a_2,b_1,b_2\in\{0,1,2\}$, $X\ket{s}=\ket{s+1~\text{mod}~3}$ and $Z\ket{s}=[\exp(2\pi{i}/3)]^s\ket{s}$, with $\ket{s}\in\{\ket{0},\ket{1},\ket{2}\}$. According to Eq.~(\ref{gp}), one can see that $\mathcal {N}_{ab}$ has $81$ possible values in total: $1$ error-free operator and $80$ error operators. The error-free operator is given by $a_1=a_2=b_1=b_2=0$ and it is in fact the identity operator acting on three-level systems $a$ and $b$. Because the depolarizing noise model is symmetric, these $80$ error operators are equally likely.

Usually, the initial state of a computation is chosen to be a very easily prepared state. Thus, the fidelity of the initial state is very high. So, we can think of the initial state of the computation as a pure state residing in the logical space $\mathscr{L}=\{\ket{0},\ket{1}\}^{\otimes{n}}$, and we denote this initial state by $\ket{\Phi_0}$. After the action of the nonadiabatic holonomic gates, the final state of the computation can be written as
\begin{eqnarray}
\mathcal {G}_{\text{after}}\cdot\mathcal {G}^\prime_{k}\cdot\mathcal {G}_{\text{before}}(\rho)
=\mathcal {G}_{\text{after}}\cdot\mathcal {N}_{ab}\cdot\mathcal {G}_{ab}
\cdot\mathcal {G}_{\text{before}}(\rho).
\label{state}
\end{eqnarray}
where $\mathcal {G}_{\text{before}}=\mathcal {G}_{k-1}\cdots\mathcal {G}_2\cdot\mathcal {G}_1$ and $\mathcal {G}_{\text{after}}=\mathcal {G}_{m}\cdots\mathcal {G}_{k+2}\cdot\mathcal {G}_{k+1}$ respectively represent the gates performed before and after $\mathcal {G}^\prime_{k}$, and $\rho=\ket{\Phi_0}\bra{\Phi_0}$.

We first consider the action of $\mathcal {G}_{\text{before}}$ and $\mathcal {G}_{ab}$ on $\rho$. Since the gates $\mathcal {G}_{\text{before}}$ and $\mathcal {G}_{ab}$ are ideal, $\mathcal {G}_{ab}\cdot\mathcal {G}_{\text{before}}(\rho)$ is a pure state residing in the logical space. Without loss of generality, this pure state can be written as
\begin{eqnarray}
\ket{\Phi}&=&\sum_{l_1l_2l_3l_4}\alpha_{l_1}\ket{l_1}\ket{0}_a\ket{0}_b
+\beta_{l_2}\ket{l_2}\ket{0}_a\ket{1}_b \nonumber \\
&&+\gamma_{l_3}\ket{l_3}\ket{1}_a\ket{0}_b
+\delta_{l_4}\ket{l_4}\ket{1}_a\ket{1}_b. \label{phi}
\end{eqnarray}
where $\alpha_{l_1}$, $\beta_{l_2}$, $\gamma_{l_3}$, $\delta_{l_4}$ are normalization coefficients, while $\ket{l_1}$, $\ket{l_2}$, $\ket{l_3}$, $\ket{l_4}$ are the states of the $n$ three-level systems except for $a$ and $b$, with $l_1$, $l_2$, $l_3$, $l_4$ being bit strings consisting of $0$ and $1$.

We next consider the action of $\mathcal {N}_{ab}$ on $\mathcal {G}_{ab}\cdot\mathcal {G}_{\text{before}}(\rho)$. Recall that $\mathcal {N}_{ab}$ has $81$ possible values: $1$ error-free operator and $80$ error operators. Before proceeding further, we divide these $80$ error operators into four subsets: $S_1$, $S_2$, $S_3$ and $S_4$. The subset $S_1$ contains the following $36$ error operators
\begin{eqnarray}
&(X)^{1}(Z)^{a_2}\otimes(X)^{1}(Z)^{b_2},& \nonumber \\
&(X)^{1}(Z)^{a_2}\otimes(X)^{2}(Z)^{b_2},& \nonumber \\
&(X)^{2}(Z)^{a_2}\otimes(X)^{1}(Z)^{b_2},& \nonumber \\
&(X)^{2}(Z)^{a_2}\otimes(X)^{2}(Z)^{b_2},&
\end{eqnarray}
where $a_2,b_2\in\{0,1,2\}$. The subset $S_2$ contains the following $18$ error operators
\begin{eqnarray}
&(X)^{1}(Z)^{a_2}\otimes(X)^{0}(Z)^{b_2},& \nonumber \\
&(X)^{2}(Z)^{a_2}\otimes(X)^{0}(Z)^{b_2}.&
\end{eqnarray}
The subset $S_3$ contains the following $18$ error operators
\begin{eqnarray}
&(X)^{0}(Z)^{a_2}\otimes(X)^{1}(Z)^{b_2},& \nonumber \\
&(X)^{0}(Z)^{a_2}\otimes(X)^{2}(Z)^{b_2}.&
\end{eqnarray}
For subset $S_4$, it contains all the rest of error operators not contained in subsets $S_1$, $S_2$ and $S_3$. That is, the subset $S_4$ contains the following $9$ error operators
\begin{eqnarray}
(Z)^{a_2}\otimes(Z)^{b_2}.
\end{eqnarray}

Consider the case where one of the error operators in subset $S_1$ occurs, and without loss of generality, we assume this error operator is $(X)^{1}(Z)^{a_2}\otimes(X)^{1}(Z)^{b_2}$, that is, $\mathcal {N}_{ab}=(X)^{1}(Z)^{a_2}\otimes(X)^{1}(Z)^{b_2}$. Note that here $a_2$ and $b_2$ are some fixed numbers. In this case, the action of $\mathcal {N}_{ab}$ on $\mathcal {G}_{ab}\cdot\mathcal {G}_{\text{before}}(\rho)$, i.e., $\mathcal {N}_{ab}\cdot\mathcal {G}_{ab}\cdot\mathcal {G}_{\text{before}}(\rho)$, is equivalent to $(X)^{1}(Z)^{a_2}\otimes(X)^{1}(Z)^{b_2}\ket{\Phi}$. By calculation, one can get that $(X)^{1}(Z)^{a_2}\otimes(X)^{1}(Z)^{b_2}\ket{\Phi}$ reads
\begin{eqnarray}
\ket{\Phi_{1a_21b_2}}&=&\sum_{l_1l_2l_3l_4}\alpha_{l_1}\ket{l_1}\ket{1}_a\ket{1}_b
+\beta_{l_2}e^{i\frac{2\pi}{3}b_2}\ket{l_2}\ket{1}_a\ket{2}_b \nonumber \\
&&+\gamma_{l_3}e^{i\frac{2\pi}{3}a_2}\ket{l_3}\ket{2}_a\ket{1}_b \nonumber \\
&&+\delta_{l_4}e^{i\frac{2\pi}{3}(a_2+b_2)}\ket{l_4}\ket{2}_a\ket{2}_b.
\end{eqnarray}
From the above equation, one can see that while the first component $\sum_{l_1l_2l_3l_4}\alpha_{l_1}\ket{l_1}\ket{1}_a\ket{1}_b$ resides in the logical space $\mathscr{L}$, the left three components $\sum_{l_1l_2l_3l_4}\beta_{l_2}e^{i\frac{2\pi}{3}b_2}\ket{l_2}\ket{1}_a\ket{2}_b +\gamma_{l_3}e^{i\frac{2\pi}{3}a_2}\ket{l_3}\ket{2}_a\ket{1}_b
+\delta_{l_4}e^{i\frac{2\pi}{3}(a_2+b_2)}\ket{l_4}\ket{2}_a\ket{2}_b$ reside in the subspace $\mathscr{L}_\perp$, i.e., the subspace orthogonal to $\mathscr{L}$.

In the above, we have discussed the action of $\mathcal {G}_{\text{before}}$, $\mathcal {G}_{ab}$ and $\mathcal {N}_{ab}$ on $\rho$, i.e., $\mathcal {N}_{ab}\cdot\mathcal {G}_{ab}
\cdot\mathcal {G}_{\text{before}}(\rho)$, where $\mathcal {N}_{ab}$ is assumed to have the value of the error operator $(X)^{1}(Z)^{a_2}\otimes(X)^{1}(Z)^{b_2}$. We then consider the action of $\mathcal {G}_{\text{after}}$, that is, $\mathcal {G}_{\text{after}}\cdot\mathcal {N}_{ab}\cdot\mathcal {G}_{ab}
\cdot\mathcal {G}_{\text{before}}(\rho)$. Specifically, after the action of $\mathcal {G}_{\text{after}}$, the state $\ket{\Phi_{1a_21b_2}}$ turns into
\begin{eqnarray}
\ket{\Phi_{1a_21b_2}}_f&=&\sum_{l_1l_2l_3l_4}\alpha_{l_1}\mathcal {G}_{\text{after}}\ket{l_1}\ket{1}_a\ket{1}_b \nonumber \\
&&+\beta_{l_2}e^{i\frac{2\pi}{3}b_2}\mathcal {G}_{\text{after}}\ket{l_2}\ket{1}_a\ket{2}_b \nonumber \\
&&+\gamma_{l_3}e^{i\frac{2\pi}{3}a_2}\mathcal {G}_{\text{after}}\ket{l_3}\ket{2}_a\ket{1}_b \nonumber \\
&&+\delta_{l_4}e^{i\frac{2\pi}{3}(a_2+b_2)}\mathcal {G}_{\text{after}}\ket{l_4}\ket{2}_a\ket{2}_b.
\end{eqnarray}
It is known that the gates $\mathcal {G}_{\text{after}}$ are ideal: $\mathcal {G}_{\text{after}}$ transform states in the logical space $\mathscr{L}$ to state in the logical space $\mathscr{L}$, and transform states in the subspace $\mathscr{L}_\perp$ to states in the subspace $\mathscr{L}_\perp$. Thus, after the action of $\mathcal {G}_{\text{after}}$, the first component still resides in the logical space $\mathscr{L}$, while the left three components still reside in the subspace $\mathscr{L}_\perp$.

We now analyse to what extent one can get benefit from using the rescaling method, i.e., Eq.~(\ref{er}), under the condition that $\mathcal {N}_{ab}=(X)^{1}(Z)^{a_2}\otimes(X)^{1}(Z)^{b_2}$. In this case, using the rescaling method is equivalent to ruling out the components of $\ket{\Phi_{1a_21b_2}}_f$ residing in the subspace $\mathscr{L}_\perp$. Note that it is the error operator $(X)^{1}(Z)^{a_2}\otimes(X)^{1}(Z)^{b_2}$ that causes the appearance of the components of $\ket{\Phi_{1a_21b_2}}_f$ residing in the subspace $\mathscr{L}_\perp$. Thus, ruling out the components of $\ket{\Phi_{1a_21b_2}}_f$ residing in the subspace $\mathscr{L}_\perp$ is equivalent to ruling out the error operator $(X)^{1}(Z)^{a_2}\otimes(X)^{1}(Z)^{b_2}$. By calculation, the probability of ruling out the error operator $(X)^{1}(Z)^{a_2}\otimes(X)^{1}(Z)^{b_2}$ reads
\begin{eqnarray}
P(1a_21b_2)&=&\bra{\Phi_{1a_21b_2}}_f(I-\hat{P})\ket{\Phi_{1a_21b_2}}_f \nonumber \\
&=&\bra{\Phi_{1a_21b_2}}(I-\hat{P})\ket{\Phi_{1a_21b_2}} \nonumber \\
&=&\sum_{l_2l_3l_4}\mid\beta_{l_2}\mid^2+\mid\gamma_{l_3}\mid^2+\mid\delta_{l_4}\mid^2,
\end{eqnarray}
where $I$ is the identity operator acting on the whole Hilbert space $\mathscr{H}$. Note that the above probability is a conditional probability and the condition is that $\mathcal {N}_{ab}$ is assumed to be the error operator $(X)^{1}(Z)^{a_2}\otimes(X)^{1}(Z)^{b_2}$. That is, under the condition of the error operator $(X)^{1}(Z)^{a_2}\otimes(X)^{1}(Z)^{b_2}$ occurring, with probability $P(1a_21b_2)$ a measurement yields an eigenstate $\ket{j}$ which does not belong to the logical subspace.

With a similar discussion, we can get the conditional probabilities $P(1a_22b_2)$, $P(2a_21b_2)$, and $P(2a_22b_2)$ that respectively describe the possibilities of ruling out the error operators $(X)^{1}(Z)^{a_2}\otimes(X)^{2}(Z)^{b_2}$,
$(X)^{2}(Z)^{a_2}\otimes(X)^{1}(Z)^{b_2}$ and $(X)^{2}(Z)^{a_2}\otimes(X)^{2}(Z)^{b_2}$ in the subset $S_1$. Specifically, these conditional probabilities read
\begin{eqnarray}
P(1a_22b_2)&=&\sum_{l_1l_3l_4}\mid\alpha_{l_1}\mid^2
+\mid\gamma_{l_3}\mid^2+\mid\delta_{l_4}\mid^2, \nonumber \\
P(2a_21b_2)&=&\sum_{l_1l_2l_4}\mid\alpha_{l_1}\mid^2
+\mid\beta_{l_2}\mid^2+\mid\delta_{l_4}\mid^2, \nonumber \\
P(2a_22b_2)&=&\sum_{l_1l_2l_3}\mid\alpha_{l_1}\mid^2
+\mid\beta_{l_2}\mid^2+\mid\gamma_{l_3}\mid^2.
\end{eqnarray}

With a similar discussion, we can also get the conditional probabilities corresponding to the error operators in the subsets $S_2$, $S_3$ and $S_4$. For example, consider the case where the error operator $(X)^{1}(Z)^{a_2}\otimes(X)^{0}(Z)^{b_2}$ occurs, i.e., $\mathcal {N}_{ab}=E_{1a_20b_2}$. Then after the action of the operator, the state $\ket{\Phi}$ turns into
\begin{eqnarray}
\ket{\Phi_{1a_20b_2}}&=&\sum_{l_1l_2l_3l_4}\alpha_{l_1}\ket{l_1}\ket{1}_a\ket{0}_b
+\beta_{l_2}e^{i\frac{2\pi}{3}b_2}\ket{l_2}\ket{1}_a\ket{1}_b \nonumber \\
&&+\gamma_{l_3}e^{i\frac{2\pi}{3}a_2}\ket{l_3}\ket{2}_a\ket{0}_b \nonumber \\
&&+\delta_{l_4}e^{i\frac{2\pi}{3}(a_2+b_2)}\ket{l_4}\ket{2}_a\ket{1}_b.
\end{eqnarray}
Then after the action of $\mathcal {G}_{\text{after}}$, the above state turns into
\begin{eqnarray}
\ket{\Phi_{1a_20b_2}}_f&=&\sum_{l_1l_2l_3l_4}\alpha_{l_1}\mathcal {G}_{\text{after}}\ket{l_1}\ket{1}_a\ket{0}_b \nonumber \\
&&+\beta_{l_2}e^{i\frac{2\pi}{3}b_2}\mathcal {G}_{\text{after}}\ket{l_2}\ket{1}_a\ket{1}_b \nonumber \\
&&+\gamma_{l_3}e^{i\frac{2\pi}{3}a_2}\mathcal {G}_{\text{after}}\ket{l_3}\ket{2}_a\ket{0}_b \nonumber \\
&&+\delta_{l_4}e^{i\frac{2\pi}{3}(a_2+b_2)}\mathcal {G}_{\text{after}}\ket{l_4}\ket{2}_a\ket{1}_b.
\end{eqnarray}
According to the above equation, we can get that the conditional probability corresponding to the error operator $(X)^{1}(Z)^{a_2}\otimes(X)^{0}(Z)^{b_2}$ reads
\begin{eqnarray}
P(1a_20b_2)&=&\bra{\Phi_{1a_20b_2}}_f(I-\hat{P})\ket{\Phi_{1a_20b_2}}_f \nonumber \\
&=&\sum_{l_3l_4}\mid\gamma_{l_3}\mid^2+\mid\delta_{l_4}\mid^2.
\end{eqnarray}
To sum up, the other conditional probabilities can also be got similarly and they can be written as
\begin{eqnarray}
P(2a_20b_2)&=&\sum_{l_1l_2}\mid\alpha_{l_1}\mid^2
+\mid\beta_{l_2}\mid^2, \nonumber \\
P(0a_21b_2)&=&\sum_{l_2l_4}\mid\beta_{l_2}\mid^2+\mid\delta_{l_4}\mid^2, \nonumber \\
P(0a_22b_2)&=&\sum_{l_1l_3}\mid\alpha_{l_1}\mid^2+\mid\gamma_{l_3}\mid^2.
\end{eqnarray}
For the error operators in the subset $S_4$, they do not cause the logical information to leak from the logical space $\mathscr{L}=\{\ket{0},\ket{1}\}^{\otimes{n}}$ because these error operators are formed by using only the generalized Pauli operator $Z$. So, the corresponding conditional probabilities have the value of zero.

We have got the conditional probabilities corresponding to each error operator. And we know that the depolarizing noise model is symmetric and therefore these error operators are equally likely. Using the above information, we can get the probability ruling out the depolarizing noise and it reads
\begin{eqnarray}
[N(S_1)+N(S_2)+N(S_3)]/80=56.25\%,
\end{eqnarray}
where $N(S_1)=\sum_{a_2b_2}P(1a_21b_2)+P(1a_22b_2)
+P(2a_21b_2)+P(2a_22b_2)=27$ is the sum of the conditional probabilities corresponding to the error operators in the subset $S_1$,
$N(S_2)=\sum_{a_2b_2}P(1a_20b_2)+P(2a_20b_2)=9$ is the sum of the conditional probabilities corresponding to the error operators in the subset $S_2$ and
$N(S_3)=\sum_{a_2b_2}P(0a_21b_2)+P(0a_22b_2)=9$ is the sum of the conditional probabilities corresponding to the error operators in the subset $S_3$. So, $56.25\%$ of the computation errors can be reduced when using the rescaling method to estimate the average value of the observable. While we assume that the depolarizing noise model is symmetric in the above, our method can also be applicable in the asymmetric case. Note that the error operators are formed by the generalized Pauli operators $X$ and $Z$, and $X$ is the reason for the logical information to leak out the logical space. Thus, if $X$ occurs with high probability and $Z$ occurs with low probability, the efficiency of our method will be increased. But if $X$ occurs with low probability and $Z$ occurs with high probability, the efficiency of our method will be decreased.

\section{Conclusion}

In conclusion, we put forward a way to estimate the average value of an observable in nonadiabatic quantum computation. The specific procedure is to perform a measurement with respect to the observable and then rescale the measurement results so that one can get a better estimation of the average value of the observable. Our way is based on the fact that while the support of the ideal final state of nonadiabatic holonomic quantum computation is a subspace of the logical subspace, the support of the noisy final state can be the whole Hilbert space. Thus projecting the noisy final state onto the logical space can increase the weight of the ideal final state, making the estimation of the average value more accurate. We use the depolarizing noise model, which is a widely adopted noise model in quantum computation, to specifically analyse to what extent one can benefit from using the rescaling method, and we find that $56.25\%$ of the computation errors can be reduced when assuming that one gate in the computation is erroneous. While our method is illustrated with $\Lambda$ system based nonadiabatic holonomic quantum computation, its application may be generalized to other quantum computation paradigms. For a quantum system used to build a qubit, it usually has many levels and two of these levels are chosen to encode the logical information. If the logical information can leak out to other levels when the quantum system experiencing inaccuracy evolutions, the logical space cannot be seen as the whole Hilbert space and our method is applicable.

\section*{Acknowledgments}
The authors acknowledge support from the National Natural Science Foundation of China
through Grant No. 12174224.

\end{document}